

\input amstex
\documentstyle{amsppt}
\magnification=1200
\parskip 10pt
\pagewidth{5.4in}
\NoBlackBoxes

\topmatter
\title
On the fundamental group of\\
the space of harmonic $2$-spheres in the $n$-sphere
\endtitle

\leftheadtext{On the fundamental group}
\rightheadtext{On the fundamental group}

\author
Mikio Furuta, Martin A. Guest, Motoko Kotani,
\\ and Yoshihiro Ohnita
\endauthor
\endtopmatter

\redefine\Bbb{\bold}
\define\C{\Bbb C}
\define\R{\Bbb R}
\define\Z{\Bbb Z}
\redefine\H{\Bbb H}
\define\Cm{\C^{2m+1}}

\define\al{\alpha}
\define\be{\beta}

\define\la{\lambda}
\define\ep{\varepsilon}

\redefine\deg{\operatorname {deg}}
\redefine\diag{\operatorname {diag}}
\redefine\dim{\operatorname {dim}}

\redefine\exp{\operatorname{exp}}
\define\Harm{\operatorname{Harm}}
\define\HH{\operatorname{HH}}
\define\Hol{\operatorname{Hol}}

\define\ddz{\frac{\partial}{\partial z}}

\redefine\sec{C^\infty}

\define\SS{S^{2m}}
\define\PP{{\bold RP}^{2m}}
\define\SSS{S^{2m+1}}

\define\CP{{\bold CP}^3}

\define\hp{\Harm_d(\PP)}
\define\hs{\Harm_d(\SS)}

\define\HS{\Harm_d(\SSS)}
\define\hz{\HH_d(Z_m)}

\define\hpf{\Harm_d^{full}(\PP)}
\define\hsf{\Harm_d^{full}(\SS)}
\define\hzf{\HH_d^{full}(Z_m)}
\define\hcp{\HH_d(\CP)}

\define\hpn{\Harm_d^{non}(\PP)}
\define\hsn{\Harm_d^{non}(\SS)}
\define\hzn{\HH_d^{non}(Z_m)}

\define\phs{\Harm_d^+(\SS)}
\define\PHS{\Harm_d^+(\SSS)}
\define\mhs{\Harm_d^-(\SS)}
\define\pmhs{\Harm_d^{\pm}(\SS)}

\define\ff{f = [f_0 ; f_1 ; f_2 ; f_3]}
\define\ffq{f = [qg_0 ; f_1 ; qg_2 ; f_3]}
\define\G{g_0g_2^\prime   - g_0^\prime  g_2}
\define\A{\al_1, \dots, \al_k}
\define\BT{\be_1, \dots, \be_N}
\define\F{\pmatrix
f_1\\
f_3\\
\endpmatrix
=
\frac{q}{\G}
\pmatrix \format \r& \quad\r \\
      g^\prime  _2  &  -g_2\\
     -g^\prime  _0  &   g_0\\
\endpmatrix
\pmatrix
h \\
h^\prime /2 \\
\endpmatrix}

\document

\baselineskip=14pt

\heading
Introduction
\endheading

Harmonic maps $\varphi:S^2\longrightarrow S^n$ are of interest from the
point
of view of differential geometry (as branched minimal immersions), and
from the point of view of gauge theory (as classical solutions of the $S^n$
non-linear sigma model). From the work of Chern, Calabi, Barbosa and
others, there is now a complete description of such harmonic maps in terms
of holomorphic data. However, very little is known about the global
properties of the space $\Harm(S^2,S^n)$ of all harmonic maps. Since the
analogous object in Yang-Mills theory --- the space of solutions to the
Yang-Mills equations --- has proved to be of great significance, the space of
harmonic maps is also of some interest. In this paper we shall give an
approach to the study of the topology of $\Harm(S^2,S^n)$.

It is known that the connected components of $\Harm(S^2,S^n)$ are indexed
by the possible energy levels $4\pi d$ of harmonic maps if $n\ge 3$,
where $d$ is a
non-negative integer. This was proved recently by Kotani (\cite{Ko}),
generalizing earlier results of Verdier (\cite{Ve1}, \cite{Ve2}, \cite{Ve3})
and Loo (\cite{Lo}). Our method gives a more conceptual proof of this (see
\cite{GO}), but our main application in this paper is to compute the
fundamental group.

\proclaim{Theorem} Let $\Harm_d(S^2,S^n)$ be the space of
harmonic maps of energy $4\pi d$, with $d>0$. Then
$$
\pi_1\Harm_d(S^2,S^n)=
\left\{
\aligned
&0\quad       \text{if}\ n\ge 4, d\ne 2\\
&\Z/2\Z\quad  \text{if}\ n\ge 4, d=2\\
&\Z/2d\Z\quad \text{if}\ n=3
\endaligned
\right.
$$
If $n=2$, $\Harm_d(S^2,S^n)$ has two connected components, each of which
has fundamental group $\Z/2d\Z$.
\endproclaim

\noindent We shall obtain from this the fundamental group of the
corresponding space $\Harm_d(S^2,\R P^n)$ (see corollary 3.10).

Our method is based on Calabi's description of harmonic maps in terms of
horizontal holomorphic maps into the appropriate twistor space, which we
summarize in \S 1.
In \S 2 we apply the method of group actions and
Morse-theoretic deformations of harmonic maps of \cite{GO} to compute the
fundamental group of the space of horizontal holomorphic maps. We combine
these results in \S 3 to obtain the fundamental group of the space of
harmonic maps. Finally, in \S 4, we give a
general construction which clarifies the geometrical significance of the
argument of \S 2, and which suggests that generalizations of the
theorem may be possible.

 $${}$$
\heading
\S 1. The space of harmonic $2$-spheres
\endheading

Let $S^n = \{ x \in \R^{n+1} \vert <x,x>\,= 1 \}$ be the $n$-dimensional
unit sphere, where $<~ ,~ >$ denotes the standard inner product of
$\R^{n+1}$. Denote by $(~ ,~ )$ the complex bilinear extension of $<~ , ~>$ to
$\C^{n+1}$. The Hermitian inner product $\ll~ , ~\gg$ on $\C^{n+1}$ is
defined by $\ll{x},{y}\gg = (x,\bar{y})$. The symbol $\perp$ will denote
orthogonal complement with respect to $\ll~,~\gg$.

Let $\Harm(S^{n})$ denote the space of all harmonic maps of $S^2$ into
$S^{n}$; we define $\Harm(\R P^n)$ similarly. The natural double covering
$p:S^n \longrightarrow \R P^n$ induces a double covering
$$
P:\Harm(S^n) \longrightarrow \Harm(\R P^n).
$$

First we recall Calabi's construction of harmonic maps of $S^2$ into $\R
P^n$ (\cite{Ca1},\cite{Ca2}, \cite{Ba}).
If a harmonic map $\varphi:S^2 \longrightarrow \R P^n$
is full, i.e. if the image of $\varphi$ is not contained in any proper
linear subspace of $\R P^n$, then $n = 2m$ for some integer $m$
(\cite{Ca1}).
The twistor space $Z_m = Z_m(\Cm)$ over $\PP$ is defined to be the
space of isotropic $m$-dimensional complex subspaces of $\Cm$. (A
complex subspace $W$ of $\Cm$ is {\it isotropic}
if $(w,w) = 0$ for each $w \in W$, or equivalently, if $W\perp\overline W$
where $\overline{W} = \{ \bar{w} \vert w \in W \}$.) Thus  $Z_m$ is a
complex submanifold of the Grassmannian $Gr_m(\Cm)$ of $m$-dimensional
complex subspaces of $\Cm$. The twistor fibration $\pi:Z_m
\longrightarrow \PP$ is defined as follows. Let $W \in Z_m$. Then we have
a decomposition $\Cm = L\oplus W\oplus \overline{W}$
where $L = (W \oplus \overline{W})^{\perp}$.
The complex line $L$ is real in the sense that $L = \bar{L}$.
Thus it has a real form $L_{\R} = \{ u \in L \vert u = \bar{u} \}
\subseteq\R^{2m+1}$. Define $\pi(W) = L_{\R} \in \PP$. The map
$\pi:Z_m \longrightarrow \PP$ is a fibre bundle with fibre
 ${O_{2m}}/{U_{m}}$. The twistor space $Z_m$ is equipped
with the K\"ahler metric induced from $Gr_m(\Cm)$, with respect to which
$\pi$ is a Riemannian submersion.

For an $(m-r)$-dimensional isotropic complex subspace $U$ of $\Cm$, we
define a subspace $Z_r^U$ of $Z_m$ by:
$$
Z_r^U =  \{ W'\oplus U \in Z_m \  \vert\ W'
\in Z_r((U \oplus \overline{U})^{\perp}) \}\cong Z_r.
$$
A map $\Phi:S^2\longrightarrow Z_m$ is said to be {\it full} if its image is
not contained in  $Z_r^U$ for any  $U$ with $r < m$. The map is said to be
{\it horizontal} with respect to $\pi$ if, for each $z\in S^2$, the image of
$(d\Phi)_z$ is orthogonal to the tangent space to the fibre of $\pi$.

The main results of \cite{Ca1},\cite{Ca2},\cite{Ba} are:

\proclaim{Theorem 1.1}
\roster
\item If $\Phi:S^2 \longrightarrow Z_m$ is a horizontal holomorphic map,
then the map
\newline
$\varphi = \pi \circ \Phi: S^2 \longrightarrow \PP$ is harmonic.
\item If $\varphi:S^2 \longrightarrow \PP$ is a full harmonic map, then
there exists a unique full horizontal holomorphic map $\Phi:S^2
\longrightarrow
Z_m$ with $\pi \circ \Phi = \varphi$.
\item For each non-constant $\Phi:S^2 \longrightarrow Z_m$, define the
positive integer $d = \deg\Phi$
by using the positive generator of $H^2(Z_m,\Z) = \Z$.
(The integer $d$ is called the {\it twistor degree} of $\Phi$.)
Then the energy of $\varphi$ in (1) is given by $E(\varphi) = 4\pi d$.  If
$\varphi: S^2\longrightarrow\PP$ is a full harmonic map, then $d\ge
m(m+1)/2$.\qed
\endroster
\endproclaim

In view of this, it is natural to introduce the following terminology:

\proclaim{Definition 1.2}
\roster
\item
Let $\hp$ be the space of all harmonic maps of $S^2$
into $\PP$ of energy $4\pi d$. Let $\hpf$ and $\hpn$ be the subspaces of
$\hp$ consisting of full maps and non-full maps, respectively.
\item
Similarly, we define $\hs$, $\hsf$, and $\hsn$.
\item
Let $\hz$ be the space of all horizontal holomorphic maps
of $S^2$ into $Z_m$ of degree $d$. Let $\hzf$ and $\hzn$ be the subspaces of
$\hz$ consisting of full maps and non-full maps, respectively.
\endroster\endproclaim

\noindent Thus, the twistor fibration $\pi:Z_m \longrightarrow \PP$
induces a surjective map
$$
\Pi:\hz \longrightarrow \hp,
$$
which restricts to a bijective correspondence between $\hzf$ and $\hpf$.

Next, we turn to the description of harmonic maps of $S^2$ into $S^n$.
The relations among $\hz$, $\hp$ and $\hs$ are depicted in the following
diagram:
$$
\matrix
{}&{}&{}&{}&{}\\
\hz            & = & \hzf             & \coprod & \hzn            \\
{}&{}&{}&{}&{}\\
\downarrow \scriptstyle{\Pi}  &   & \updownarrow \scriptstyle{1:1} &    &
\downarrow onto \\
{}&{}&{}&{}&{}\\
\hp            & = & \hpf             & \coprod & \hpn            \\
{}&{}&{}&{}&{}\\
\uparrow\scriptstyle{P}  &   & \uparrow\scriptstyle{2:1}
&         & \uparrow \scriptstyle{2:1}    \\
{}&{}&{}&{}&{}\\
\hs            & = & \hsf             & \coprod & \hsn          \\
{}&{}&{}&{}&{}\\
\endmatrix
$$
To clarify the relation between $\hz$ and $\hs$, we recall the definition of
the twistor fibrations $\pi_{\pm}:Z_m \longrightarrow S^{2m}$. Choose an
orientation of $\R^{2m+1}$. Let $W \in Z_m$, so that $\Cm = L \oplus W
\oplus \overline{W}$ as above.  There is an orthogonal complex structure
$J_W$ on $(W \oplus \overline{W})_{\R} \subseteq \R^{2m+1}$ such that
$W$ is the $\sqrt{-1}$-eigenspace of $J_W$.  This complex structure
defines a natural orientation of $(W \oplus \overline{W})_{\R}$.
There is a unique unit vector $x \in \SS$ such that the orientation
of $\{ x \} \oplus (W \oplus \overline{W})_{\R}$ agrees with the chosen
orientation of $\R^{2m+1}$. We define $\pi_{\pm}(W) = \pm x$.
Obviously we have $\pi = p \circ \pi_{\pm}$, i.e., the following diagram is
commutative:
$$
\matrix
 & & \\
                       & Z_m &                        \\
 & & \\
\scriptstyle{\pi_+}\swarrow     &     & \searrow \scriptstyle{\pi_-}    \\
 & & \\
  \SS           &\downarrow \scriptstyle{\pi} &           \SS  \\
 & & \\
\scriptstyle{p}\searrow &     & \swarrow\scriptstyle{p} \\
 & & \\
                       & \PP &                        \\
 & & \\
\endmatrix
$$
The twistor fibrations $\pi_{\pm}$ induce maps $\Pi_{\pm}:\hz
\longrightarrow \hs$, and we have $\Pi = P \circ \Pi_{\pm}$, i.e., the
following diagram is commutative:
$$
\matrix
 & & \\
                       & \hz &                        \\
 & & \\
\scriptstyle{\Pi_+}\swarrow     &     & \searrow \scriptstyle{\Pi_-}    \\
 & & \\
  \hs           &\downarrow \scriptstyle{\Pi} &           \hs  \\
 & & \\
\scriptstyle{P}\searrow &     & \swarrow\scriptstyle{P} \\
 & & \\
                       & \hp &                        \\
 & & \\
\endmatrix
$$
In contrast to $\Pi$ and $P$, the maps $\Pi_\pm$ are not surjective if
$m>1$, so we introduce some further notation:

\proclaim{Definition 1.3}
Let $\pmhs$ be the image of the map $\Pi_{\pm}$.
\endproclaim

For $i=0,\dots,m$, let
$$
{\Cal X}_i=\{\Phi\in\hz\ \vert\ \Phi(S^2)\subseteq Z_{m-i}^{U}\ \text{for
some}\  U\in Gr_{i}(\Cm)\}.
$$
This gives a filtration
$$
\hz=\Cal X_0\supseteq \hzn=\Cal X_1\supseteq\dots\supseteq\Cal
X_m=\emptyset
$$
of $\hz$ by closed subspaces. Similarly, let
$$
{\Cal Y}^\pm_i=\{\varphi\in\pmhs\ \vert\
\varphi(S^2)\subseteq S^{2m}\cap V
\ \text{for some}\ V\in Gr_{2m+1-2i}(\R^{2m+1})\},
$$
so that we have a filtration
$$
\pmhs=\Cal Y^{\pm}_0\supseteq\hsn=\Cal
Y^{\pm}_1\supseteq\dots\supseteq\Cal Y^{\pm}_m=\emptyset.
$$
We note that $\Cal Y_i^+=\Cal Y_i^-$ when $i>0$. Hence:

\proclaim{Proposition 1.4} We have $\hs = \phs\cup\mhs$,
\newline
with $\phs\cap\mhs=\hsn$.\qed
\endproclaim

For each $i=0,\dots,m-1$ the restriction of $\Pi_\pm$ defines a fibre
bundle
$$
\Pi_{\pm} : {\Cal X}_i-{\Cal X}_{i+1}\longrightarrow{\Cal Y}_i^{\pm}-{\Cal
 Y}_{i+1}^{\pm}
$$
with fibre $SO_{2i}/U_{i}$.

 This description of $\hs$ was obtained by Verdier
(\cite{Ve1},\cite{Ve2}), in the case $m=2$. He conjectured that $\hs$ is
path connected. This was proved in the case $m=2$ by Loo (\cite{Lo}) and by
Verdier himself (\cite{Ve3}), and
for the general case $m\ge 2$ by Kotani (\cite{Ko}):

\proclaim{Theorem 1.5} The space $\hs$ is path connected for $m\ge 2$.
\qed
\endproclaim

\noindent In fact, the method of \cite{Ko} also gives:

\proclaim{Theorem 1.6}
The spaces ${\Cal X}_i$, ${\Cal Y}^\pm_i$ are all path connected for $m\ge
2$. \qed
\endproclaim

\noindent In particular $\pmhs$ and $\hsn$ are path connected for $m\ge 2$.

Concerning $\Harm_d(S^n)$ for $n=2$ or $n$ odd, we have:

\noindent (1) $\Harm_d(S^2)$ is the disjoint union of
$\Harm_{d}^{+}(S^2)$ and $\Harm_{d}^{-}(S^2)$. Each of these is a copy of the
space $\Hol_d(S^2)$ of all holomorphic maps from $S^2$ to $S^2$ of degree
$d$. Since $\Hol_d(S^2)$ is an irreducible algebraic variety, it is connected,
so $\Harm_d(S^2)$ has two connected components.

\noindent (2) $\HS$ is also path connected for $m\ge 1$. To see this, note
that $\phs$ is path connected for $m\ge 1$ and every $SO_{2m+2}$-orbit in
$\HS$ intersects $\phs$.

\noindent (3) $\Harm(S^1)$ consists of the constant maps.
$${}$$

\heading
\S 2. The fundamental group of $\HH_d(Z_m)$
\endheading

In this section we shall investigate the space of horizontal holomorphic
maps:

\proclaim{Theorem 2.1} Let $m\ge2$ and $d>0$. Then

\noindent (1) $\pi_1\HH_d(Z_m)=0$ if $d\ne2$

\noindent (2) $\pi_1\HH_2(Z_m)=\Z/2\Z$.
\endproclaim

We begin with some elementary geometry of the space $Z_m$. For $V\in
Z_m$, let
\newline
$\pi_V:\C^{2m+1}\longrightarrow V$ denote the orthogonal
projection operator with respect to $\ll~ , ~\gg$. Then
$P=i(\pi_V-\pi_{\bar V})$ defines an element of the Lie algebra of
$SO^{\C}_{2m+1}$, i.e. of
$$
\frak{so}^{\C}_{2m+1}=\{A\in\text{End}(\C^{2m+1})\ \vert\ A^t=-
A,\text{tr}A=0\}.
$$
In fact, $P$ belongs to the Lie algebra of $SO_{2m+1}$, i.e. to the
subspace
$$
\frak{so}_{2m+1}=\{A\in\frak{so}^{\C}_{2m+1}\ \vert\ \bar A=A\}.
$$
We may identify $Z_m$ with the orbit of $P$
under the adjoint representation of $SO_{2m+1}$.

Let $L$ be an isotropic line in $\C^{2m+1}$, and set $Q=i(\pi_L-
\pi_{\bar L})$. Then $\{\exp(itQ)\}_{t\in\R}$ is a one parameter
subgroup of $SO^{\C}_{2m+1}$, and its action on $Z_m$ gives the
gradient flow of a natural Morse-Bott function $f^Q:Z_m\longrightarrow\R$
(see \S 6 of \cite{GO}). The critical points $W$ of $f^Q$ are the
fixed points of the action of the one parameter subgroup
$\{\exp(tQ)\}_{t\in\R}$ of $SO_{2m+1}$, and are given by the
condition
$$
[i(\pi_W-\pi_{\bar W}),\ i(\pi_L-\pi_{\bar L})]=0.
$$
This means that $W$ is of the form $W_1\oplus W_2\oplus W_3$, where
$W_1\subseteq L$, $W_2\subseteq\bar L$, and $W_3\subseteq(L\oplus\bar
L)^\perp$. As $W$ is
isotropic, exactly one of $W_1$ and $W_2$ must be $\{0\}$. Hence,
there are two connected critical manifolds, namely:
$$\gather
Z_m^+=\{W\in Z_m\ \vert\ L\subseteq W\subseteq\bar L^\perp\}\cong
Z_{m-1}\\
Z_m^-=\{W\in Z_m\ \vert\ \bar L\subseteq W\subseteq L^\perp\}\cong
Z_{m-1}.
\endgather
$$
The embeddings $I^\pm:Z_{m-1}\longrightarrow Z_m$, given by the
inclusions of
$Z_m^\pm$ in $Z_m$, respect the holomorphicity condition
and the horizontality condition.
In other words, a map
\newline $f:S^2\longrightarrow Z_{m-1}$ satisfies these
conditions if and only if the composition $I^\pm\circ f$ does.

We shall make use of the following subspace of $\HH_d(Z_m)$:

\proclaim{Definition 2.2} $\HH^\ast_d(Z_m)=\{f\in \HH_d(Z_m)\
\vert\ f(S^2)\cap Z_m^+=\emptyset\}.$
\endproclaim

\noindent From the preceding discussion, we can identify
$\HH_d( Z_{m-1})$ with the subspace $\HH_d(Z_m^-)$ of $\HH_d^\ast
(Z_m)$.

\proclaim{Lemma 2.3} $\HH_d(Z_{m-1})$ is a deformation retract
of $\HH_d^\ast(Z_m)$.
\endproclaim

\demo{Proof}
The flow given by the one parameter subgroup
$\{\exp(itQ)\}_{t\in\R}$ provides a deformation retraction
$W\longmapsto\lim_{t\to\infty}\exp(itQ)W$ of $Z_m-Z_m^+$ onto $Z_m^-$.
The action of $\exp(itQ)$ preserves the holomorphicity and
horizontality conditions, as the action of the whole of
$SO^{\C}_{2m+1}$ does (see \cite{GO}). Hence, for any $f\in
\HH_d^\ast(Z_m)$, we
obtain a curve $\{f_t\}_{0\le t\le\infty}$ in $\HH_d^\ast(Z_m)$ by
defining $f_t=\exp(itQ)f$. The map $f\longmapsto f_\infty$
gives the required deformation retraction onto $\HH_d(Z_{m-1})$.
\qed\enddemo

With these preliminaries out of the way, we come to the proof of
theorem 2.1.

\demo{Proof of theorem 2.1(1)} For $d\ne 2$ we shall prove

\noindent (A) that the induced map $\pi_1\HH_d^\ast
(Z_m)\longrightarrow\pi_1\HH_d(Z_m)$ is surjective (if $m\ge 2$), and

\noindent (B) that the induced map $\pi_1\HH_d^\ast
(Z_2)\longrightarrow\pi_1\HH_d(Z_2)$ is zero.

\noindent The theorem then follows by induction on $m$, as the case
$m=2$ is given directly by (A) and (B), while the inductive step
follows from (A) and lemma 2.3.

To prove (A), we must show that any loop $\{f_t\}_{0\le t\le 1}$ in
$\HH_d(Z_m)$ may be deformed into $\HH_d^\ast(Z_m)$. It suffices
to consider smooth loops. We claim that there exists a
matrix $A\in SO_{2m+1}$ such that $Af_t(z)\notin Z_m^+$, i.e.
$Af_t(z)\not\supseteq L$, for all $t\in[0,1], z\in S^2$. The required
deformation can then be constructed by taking a path from $A$ to
the identity in $SO_{2m+1}$, as $SO_{2m+1}$ is a connected group
which acts on $\HH_d(Z_m)$. Since $SO_{2m+1}$ acts transitively
on the space $Y_m$ of isotropic lines in $\C^{2m+1}$, it suffices to
find some $L^\prime\in Y_m$ such that $f_t(z)\not\supseteq
L^\prime$ for all $t,z$. Let
$$
Y^f_m=\{L^\prime\in Y_m\ \vert\ L^\prime\subseteq f_t(z)\ \text{for
some}\ t,z\}.
$$
Then our claim is that $Y^f_m\ne Y_m$.

Consider the flag manifold $X_m=\{(L^\prime,V)\in Y_m\times Z_m\
\vert\ L^\prime\subseteq V\}$. We have projection maps
$y:X_m\longrightarrow Y_m$,
$z:X_m\longrightarrow Z_m$, and
$$
Y^f_m=y(z^{-1}(f([0,1]\times S^2))).
$$
The map $z$ is a fibre bundle, with fibre $\C P^{m-1}$. The maps
$y,f$ do not increase dimension, as they are smooth+. So
we have: $\dim\, Y^f_m\le\dim\,z^{-1}(f([0,1]\times S^2))\le
\allowmathbreak (2m-2)+\dim\,f([0,1]\times S^2)\le (2m-
2)+3=2m+1$. But $\dim\, Y_m=4m-2$, so $Y_m^f$ cannot be equal to
$Y_m$ if $m\ge2$. This completes the proof of (A).

Regarding (B), it suffices (because of lemma 2.3) to show that the
map $\pi_1\HH_d(Z_2^-)\longrightarrow\pi_1\HH_d(Z_2)$ is zero. The
fundamental group of $\HH_d(Z_2^-)\cong \Hol_d(S^2)$ is
known to be cyclic of order $2d$ (\cite{Ep}).
If points of $\Hol_d(S^2)$ are represented in the usual way as
rational functions $p/q$ (where $p$ and $q$ are coprime polynomials
with maximal degree $d$), then a loop $\al:S^1\longrightarrow \Hol_d(S^2)$
which generates $\pi_1\Hol_d(S^2)$ may be described as follows.
Let $\al(\la)=p_\la/q_\la$, for any monic $p_\la,q_\la$ such that
the map $\la\longmapsto R(p_\la,q_\la)$ (the resultant of $p_\la$ and
$q_\la$) has winding number $\pm1$. Less formally, this means that
\lq\lq a simple root of $p_1$ is moved once around a simple root of
$q_1$\rq\rq. We must show that the loop $\al$ is contractible in
$\HH_d(Z_2)$.

First we recall the well known identification of $Z_2$  with $\C
P^3$ (c.f. \cite{Br}), which will enable us to work with this more familiar
space. We have $Z_2\cong SO_5/U_2\cong Sp_2/S^1\times
Sp_1$ and $\C P^3\cong SU_4/S(U_1\times U_3)$. The identification
is induced by the inclusion homomorphism $Sp_2\longrightarrow SU_4$,
where $Sp_2$ is considered as a subgroup of $SU_4$ via the identification
of right $\H$-modules $\H^2\longrightarrow\C^4$,
$(a+jb,c+jd)\longmapsto(a,b,c,d)$. We also have an inclusion
$Sp^{\C}_2\longrightarrow SU^{\C}_4=Sl_4\C$, whereby
$Sp^{\C}_2$ is identified with the subgroup $\{X\in Sl_4\C\ \vert\
X^\ast j=jX^{-1}\}$ (here, $j$ denotes the operator given by right
multiplication of the quaternion $j$ on $\C^4$). The twistor fibration
$\pi_+:Z_2\longrightarrow S^4$ becomes the natural map $Sp_2/S^1\times
Sp_1\longrightarrow
Sp_2/Sp_1\times Sp_1$, and a holomorphic map $f:S^2\longrightarrow\C
P^3$ is
horizontal with respect to $\pi_+$ if and only if
$$
\ddz\sec(\underline{f})\perp\sec(j\underline{f}).
$$
Here $\underline{f}$ denotes the subbundle of the trivial bundle
$S^2\times \C^4$ corresponding to the map $f$.
If $f$ is written using homogeneous coordinates as
$f=[f_0;f_1;f_2;f_3]$, where $f_0,f_1,f_2,f_3$ are coprime
polynomials, then this horizontality condition is
$$
f_0f^\prime_1-f^\prime_0f_1 + f_2f^\prime_3-f^\prime_2f_3=0.
$$
The action of $Sp^{\C}_2$ on $\C P^3$ preserves this horizontality
condition (but the action of $Sl_4\C$ does not, in general).

We may identify $\C P^3$ with the adjoint orbit of $P=\diag(i,0)$ in
$\frak{sp}_2$. Let $Q=\diag(i,i)$ in $\frak{sp}_2$.
The one parameter subgroup $\{\exp(itQ)\}_{t\in\R}$ of
$Sp^{\C}_2$, which corresponds to the one parameter
subgroup $\{\diag(e^{t},e^{-t},e^{t},e^{-t})\}_{t\in\R}$ of $Sl_4\C$,
generates the gradient flow of a Morse-Bott function on $\C P^3$.
(Up to a constant, the function is given explicitly by
$[z_0;z_1;z_2;z_3]\longmapsto(\vert z_1\vert^2+\vert
z_3\vert^2)/\sum\vert z_i\vert^2$.)
The connected critical manifolds of this function are:
$$\gather
Z^+=\{[z_0;z_1;z_2;z_3]\in\C P^3\ \vert\ z_0=z_2=0\}\cong S^2\\
Z^-=\{[z_0;z_1;z_2;z_3]\in\C P^3\ \vert\ z_1=z_3=0\}\cong S^2.
\endgather
$$
 The deformation retraction $D:\HH^\ast_d(\C
P^3)\longrightarrow\Hol_d(S^2)$ is given by
$[f_0;f_1;f_2;f_3]\longmapsto[f_0;f_2]$.

Now we return to the problem of constructing a null-homotopy of the
loop $\al$ in $\HH_d(\C P^3)$. We shall construct a loop $\tilde\al$ which is
contractible in $\HH_d(\C P^3)$, but which is homotopic to $\al$.

\proclaim{Lemma 2.4}
If $d \ne 2$, then there is an element $\ff \in \hcp$ which satisfies:
\roster
\item $f_0$ and $f_2$ have precisely one common root, at 0.
\item $f_1(0) \ne 0$.
\endroster
(If $\deg\ f_i <d$ here, then we consider $\infty$ to be a root of $f_i$
with multiplicity $d-\deg\ f_i$.)
\endproclaim

\demo{Proof} If $d=1$ we may take $f(z)=[z;1;z;-1]$. If $d>2$
it is straightforward to check that
$$
f(z)=[(d-2)z^d+z^{d-1}+z;
    z^d+dz^{d-1}+1;
   (d-2)z^d-z^{d-1}-z;
    z^d-dz^{d-1}+1]
$$
 satisfies conditions (1) and (2). (This is a variant of an example given in
\cite{Lo}. Its geometrical significance will be discussed later, in \S 4.)
\qed\enddemo

\noindent Let $A:S^1\longrightarrow Sp_2$ be a loop in $Sp_2$. Define
$\tilde\al(\la)=A(\la)f$. Then $\tilde\al$ is a contractible loop in
$\HH_d(\C P^3)$, as $Sp_2$ is simply connected and acts on
$\HH_d(\C P^3)$. We claim that $A$ may be chosen so that
$\tilde\al\simeq\al$. It will suffice to find a loop $A$ such that

\noindent (i) $\tilde\al(S^1)\subseteq \HH_d^\ast(\C P^3)$, and

\noindent (ii) $D(\tilde\al)\simeq\al$.

\noindent If we write
$\tilde\al(\la)=[f_0^\la;f_1^\la;f_2^\la;f_3^\la]$, then conditions
(i),(ii) say that $f_0^\la,f_2^\la$ are coprime, and the map
$\la\longmapsto R(f_0^\la,f_2^\la)$ has winding number $\pm1$. These
conditions are satisfied by
$A(\la)=\diag(u_\la+jv_\la,1)$, where $v_\la=\ep\la$,
$u_\la=(1-\ep^2)^{\frac12}$ and $\ep$ is a small positive
number. To see this, first notice that
the roots of $f_0^\la$ are obtained by perturbing slightly the roots of $f_0$,
while the roots of $f_2^\la=f_2$ do not depend on $\la$,
in particular $0$ is a root of $f_2^\la$ for any $\la$.
Let $z_{\la,\ep}$ be the root of $f_0^\la$ which is close to the root $0$ of
$f_0$. Since $f_0^\la=f_0$ at $\ep=0$, and $f_0^{\prime}(0)\neq 0$, the
implicit function theorem implies that $z_{\la ,\ep}$
is well defined and satisfies
$z_{\la ,\ep} =O(\ep)$.
The  only contribution to the winding number comes from the
rotation of $z_{\la ,\ep}$ around $0$.  Since $f$ is of the form
$[az+O(z^2);b+O(z);f_2(z);f_3(z)]$ for some $a,b\neq 0$, we have
$$
f_0^\la(z) =  u_\la f_0(z) - \bar v_\la f_1(z)
   = (1-\ep^2)^{\frac12}f_0(z)-\ep\bar{\la}f_1(z)
   = az - b\ep\bar{\la} + O(z^2)+O(\ep z).
$$
Since $z_{\la ,\ep} = O(\ep)$, we obtain
$$
0 =  az_{\la ,\ep} - b\ep\bar{\la} + O(\ep^2),
$$
hence
$
z_{\la ,\ep} = (\ep b/a)\bar{\la} + O(\ep^2)
$
moves once around $0$.
\qed\enddemo

\demo{Proof of theorem 2.1(2)} This will be given in the next section.
\enddemo

 $${}$$
\heading
\S 3. The fundamental groups of $\hs$ and $\HS$
\endheading

{}From proposition 1.4 we have $\hs=\phs\cup\mhs$. As a first step in
computing the fundamental groups of $\hs$ and $\HS$, we shall prove:

\proclaim{Theorem 3.1} Let $m\ge 2$ and $d>0$. Then
\roster
\item $\pi_{1}\Harm^{\pm}_{d}(S^{2m}) = 0$ if $d\ne 2$
\item $\pi_{1}\Harm^{\pm}_{2}(S^{2m}) = {\bold Z}/2{\bold Z}$.
\endroster
\endproclaim

\noindent We shall make use of the following two lemmas, whose proofs are
postponed until later.

\proclaim{Lemma 3.2}
 Suppose $f:E\longrightarrow B$ is a continuous map between
locally contractible spaces such that
\roster
\item  $f^{-1}(b)$ is non-empty and connected for all $b\in B$, and

\item $f$ is  a closed map.
\endroster
Then
$f_{*}:\pi_{1}E \longrightarrow \pi_{1}B$ is surjective.
\endproclaim

\proclaim{Lemma 3.3}
The map
$\Pi_{\pm}:\HH_{d}(Z_{m}) \longrightarrow \Harm^{\pm}_{d}(S^{2m})$ is
proper.
\endproclaim

\demo {\it Proof of theorem 3.1(1)}
Recall from \S 1 that we have surjections
$\Pi_\pm:\HH_{d}(Z_{m})\longrightarrow \Harm_{d}(S^{2m})$, under which
the
inverse image of any point is of the form $SO_{2i}/U_{i}$. Since they
are described by algebraic equations, both $\hz$ and $\pmhs$ are locally
triangulable (\cite{Sa}), hence locally compact and locally contractible.
Recall that a proper continuous map to a locally compact (Hausdorff) space
is a closed map. By lemma 3.3, $\Pi_\pm$ is closed, and so
by lemma 3.2, $(\Pi_\pm)_\ast$ is surjective. In the case $d\ne 2$ we have
$\pi_{1}\HH_{d}(Z_{m})=0$ (by theorem 2.1(1)), hence
$\pi_{1}\Harm^{\pm}_{d}(S^{2m})=0$, as required.
\qed
\enddemo

\demo{Proofs of theorem 2.1(2) and theorem 3.1(2)}
For $d=2$, the argument of the proof of theorem 2.1(1) is valid except for
lemma 2.4. However, by (A) and lemma 2.3, the map
$\pi_1\Hol_2(S^2)\longrightarrow\pi_1\HH_2(\C P^3)$ is surjective, and
from the argument of (B) we see that
$2[\al]\in\pi_1\Hol_2(S^2)=\Z/4\Z$ maps to zero.
(Use $\ff =[z^2-1,z^2,z^2-1,-1] \in \HH_2(\C P^3)$; since
$f_0$ and $f_2$ have precisely two common roots, the argument of (B)
shows $2[\al]\mapsto 0$.)
Hence
$\pi_1\HH_2(\C P^3)$ is either $0$ or $\Z/2\Z$. The same induction
argument as in the proof of theorem 2.1(1) then shows that
$\pi_1\HH_2(Z_m)$ is either $0$ or $\Z/2\Z$.

To prove that $\pi_1\HH_2(Z_m)=\Z/2\Z$, we shall suppose that it is zero,
and obtain a contradiction. By theorem 1.1(3) we have
$$
\HH_2(Z_m)=\HH_2^{non}(Z_m)
$$
and
$$
\Harm_2(S^{2m})=\Harm^{\pm}_2(S^{2m})=\Harm_2^{non}(S^{2m}).
$$
If $\pi_1\HH_2(Z_m)=0$, then by lemmas 3.2 and 3.3 it follows that
$\pi_1\Harm_2(S^{2m})=0$. Now, again by theorem 1.1, the image of any
$f\in\Harm_2(S^{2m})$ must be of the form $S^{2m}\cap V$, for some
$V\in\widetilde{Gr}_3(\R^{2m+1})$ (the Grassmannian of oriented
$3$-planes in $\R^{2m+1}$). So $\Harm_2(S^{2m})$ is the total space of
a fibre bundle over $\widetilde{Gr}_3(\R^{2m+1})$ with fibre
$\Hol_{2}(S^2)$. The homotopy exact sequence now gives a contradiction, as
$\pi_{2}\widetilde{Gr}_3(\R^{2m+1})={\Z}/2{\Z}$ and
$\pi_{1}\Hol_{2}(S^2) = {\Z}/4{\Z}$. This completes the proof of theorem
2.1(2).

 The same argument also proves theorem 3.1(2). Indeed, as in the proof of
theorem 3.1(1), the map
$\Pi_\pm:HH_2(Z_m)\longrightarrow\Harm_2^\pm(S^{2m})$ induces a
surjection on fundamental groups, so $\pi_1\Harm_2^\pm(S^{2m})$ is either
$0$ or $\Z/2\Z$. By the previous paragraph, however, it cannot be $0$.
\qed
\enddemo

We can now prove our main results.

\proclaim{Theorem 3.4}
Let $m\ge 2$ and $d>0$. Then
\roster
\item $\pi_{1}\hs = 0$ if $d\ne 2$
\item $\pi_{1}\Harm_2(\SS) = {\bold Z}/2{\bold Z}$.
\endroster
\endproclaim

\demo{Proof} For two locally contractible spaces with connected
intersection, van Kampen's theorem can be applied (\cite{Ol}). We obtain
$\pi_{1}\Harm_{d}(S^{2m})=0$ for $d \ne 2$ by applying this to the union of
$\Harm^{+}_{d}(S^{2m})$ and
$\Harm^{-}_{d}(S^{2m})$, since $\pi_1\pmhs=0$ by theorem 3.1(1) and
$\hsn$ is connected by theorem 1.6. For $d=2$, the result is immediate from
theorem 3.1(2), as $\Harm_2(S^{2m})=\Harm^{\pm}_2(S^{2m})$.
\qed\enddemo
\noindent

Next we consider the case of an odd dimensional sphere.

\proclaim{Theorem 3.5}
Let $m\ge 1$ and $d>0$. Then
\roster
\item $\pi_{1}\HS = 0$ if $m\ge 2$ and $d\ne 2$
\item $\pi_{1}\Harm_2(\SSS) = {\bold Z}/2{\bold Z}$ if $m\ge 2$.
\item $\pi_{1}\Harm_d(S^3) = {\bold Z}/2d{\bold Z}$.
\endroster
\endproclaim

\noindent We shall in fact show:

\proclaim{Proposition 3.6} Let $m\ge 1$ and $d>0$. Then the natural
inclusion map $\SS\subset\SSS$ induces an isomorphism
$$
\pi_1\phs = \pi_1\HS.
$$
\endproclaim
\demo{Proof of theorem 3.5 assuming proposition 3.6}
For $m\ge 2$, statements (1) and (2) follow from theorem 3.1.
For $m=1$, the result follows from the fact that
$\Harm_2^+(S^2)=\Hol_d(S^2)$ and
$\pi_1\Hol_d(S^2)={\bold Z}/2d{\bold Z}$.
\qed
\enddemo

To compare $\HS$ and $\phs$, we use a space $\PHS$ which has
two equivalent definitions. One definition is that $\PHS$
is the total space of the fibre bundle over $\SSS$ whose fibre over
$v\in\SSS$ is $\Harm^+(\SSS\cap v^\perp )$. (Note that $v^\perp$ is
oriented.) The other definition is that
$\PHS$ is the subspace of $\SSS\times\HS$ which consists of
all pairs $(v,f)$ such that the image of $f$ is perpendicular to $v$ and
if the image spans $v^\perp$, then the orientation of
$v^\perp$ induced from $f$ agrees with the canonical one which comes from
the orientation of $\R^{2m+2}$ and the direction of $v$.
 Let $p_0:\PHS\longrightarrow\SSS$  be the projection to the first factor
and $p_1:\PHS\longrightarrow\HS$  that to the second factor. Then $p_0$ is
the projection for the bundle structure in the first definition.
The second definition immediately implies:

\proclaim{Lemma 3.7}
 The map $p_1$ is proper.\qed
\endproclaim

\noindent
By theorem 1.1 the image of any $f\in \HS$ must be of the form
$\SSS\cap V$, for some $V\in \widetilde{Gr}_{2i+1}(\R^{2m+2})$.
So $p_1$ is surjective. If $i=m$, then  $p_1^{-1}(f)$ is
one point. If $i<m$, then $p_1^{-1}(f)$ is
the set of unit vectors perpendicular
to $V$, hence homeomorphic to $S^{2(m-i)}$.
In particular, since if $d=2$ then $i=1$, we have:

\proclaim{Lemma 3.8} Let $m\ge 1$ and $d>0$.
\roster
\item Every fibre of $p_1$ is a non-empty connected space.
\item If $d=2$ and $m\ge 2$,
then $p_1:\Harm_2^+(\SSS)\longrightarrow\Harm_2(\SSS)$
 is a fibre bundle with fibre $S^{2(m-1)}$.
\item If $m=1$, then $p_1:\Harm_d^+(S^3)\longrightarrow\Harm_d(S^3)$
 is a homeomorphism.\qed
\endroster
\endproclaim

\demo{Proof of proposition 3.6}
{}From the homotopy exact sequence for the fibre bundle
\newline
$p_0:\PHS\longrightarrow\SSS$, we have
$\pi_1\PHS = \pi_1\phs$.
By lemmas 3.7 and 3.8 (1),
we can apply lemma 3.2 to $p_1$,
as in the proof of theorem 3.1(1), to obtain that the induced map
$\pi_1\PHS\longrightarrow\pi_1\HS$ is surjective.
If $m\ge 2$ and $d\ne 2$, then
$\pi_1\PHS =\pi_1\phs =0$ by theorem 3.1(1), hence
$\pi_1\HS=0$.
If $m=1$ or $d=2$, then, by lemma 3.8 (2),(3)
we have the homotopy exact sequence for
 $p_1:\Harm_2^+(\SSS)\longrightarrow\Harm_2(\SSS)$ which implies that
$\pi_1 \Harm_2^+(\SSS) = \pi_1 \Harm_2(\SSS)$, hence we have the
required
isomorphism.
\qed
\enddemo

\noindent Theorem 3.4 and theorem 3.5 together give the theorem stated in
the introduction. (Note that $\Harm_d(S^2)$ consists of two copies of
$\Harm_d^+(S^2)$, i.e. of $\Hol_d(S^2)$, and we have
$\pi_1\Hol_d(S^2)=\Z/2d\Z$ by \cite{Ep}.)

To determine $\pi_1\Harm_d(S^2,\R P^n)$, we use the following proposition.

\proclaim{Proposition 3.9} Let $n\ge 3$ and $d>0$. Then
the homotopy exact sequence
$$
0 \longrightarrow \pi_{1}\Harm_{d}(S^{n})
  \longrightarrow \pi_{1}\Harm_{d}({\bold R}P^{n})
  \longrightarrow {\bold Z}/2{\bold Z}
  \longrightarrow 0
$$
for the double covering $P:\Harm_{d}(S^{n}) \longrightarrow
\Harm_{d}({\bold R}P^{n})$ splits.
\endproclaim

\demo{Proof}
Let $R(\theta)\in SO_2$ be rotation through angle $\theta$
and  let $g(\theta)\in SO_{n+1}$  be the direct sum
$R(\theta)\oplus R(\theta)\oplus I_{n-3}$, where $I_{n-3}$ is the
identity matrix of dimension $n-3$. Take some $f\in\Harm_d(S^{n})$
whose image is contained in
${\R}^{4}\times(0,\dots,0)\subseteq {\bold R}^{n+1}$.
The path $\{ g(\theta)f\ | \ 0\le\theta\le \pi\}$
in $\Harm_{d}(S^{n})$ projects to a closed path $\varphi$ in
$\Harm_{d}(\R P^{n+1})$. From the construction $\varphi$ cannot be lifted to
a closed path in $\Harm_{d}(S^{n})$.
To give the splitting it suffices to show that the order of the class
of $[\varphi ] \in\pi_{1}\Harm_{d}(\R P^{n+1})$ is two.
The composite of $\varphi$ with itself is given by the projection of the
closed path $\{ g(\theta)f\ | \ 0\le\theta\le 2\pi\}$ in $\Harm_{d}(S^{n})$.
The closed path $\{ g(\theta)\ | \ 0\le\theta\le 2\pi\}$ in $SO_{n}$
is contractible from the definition of $g$, hence so is the path
$\{g(\theta)f\ \vert\ 0\le\theta\le 2\pi\}$ and its
projection.
\qed
\enddemo

\noindent Our result for $\R P^n$ is:

\proclaim {\bf Corollary 3.10} Let $d>0$. Then
$$
\pi_1\Harm_d(\R P^n)=
\left\{
\aligned
&\Z/2\Z\quad       \text{if}\ n\ge 4, d\ne 2\\
&\Z/2\Z\oplus\Z/2\Z\quad  \text{if}\ n\ge 4, d=2\\
&\Z/2d\Z\oplus\Z/2\Z\quad \text{if}\ n=3\\
&\Z/2d\Z\quad \text{if}\ n=2
\endaligned
\right.
$$
\endproclaim

\demo{Proof} For $n>2$, this follows from proposition 3.9, theorem 3.4 and
theorem 3.5. For $n=2$, $\Harm_d(\R P^2)$ may be identified with
$\Hol_d(S^2)$, and we have seen already that the fundamental group of this
space is $\Z/2d\Z$.
\qed\enddemo

Finally, we give the proofs of lemmas 3.2 and 3.3.

\demo {\it Proof of lemma 3.2}
This is an elementary lemma of general topology.
We show that for any closed path $\varphi:S^{1} \longrightarrow B$ there is
a
path
$\psi:S^{1} \longrightarrow E$
whose projection $f\circ \psi :S^{1}\longrightarrow B$ is
homotopic to $\varphi$. Take  finitely many contractible open subsets
$U_{1},\dots,U_{l}$
of $B$ such that the image of $f\circ \psi$ is contained in the union of
the $U_{i}$'s. We may assume that for a decomposition
$ S^{1} = [0, 1] \mod 1  = [0, t_{1}] \cup [t_{1}, t_{2}] \cup \dots
\cup [t_{l-1},1]$ mod $1$,  the image of $[t_{i-1},t_{i}]$ lies in $U_{i}$
($t_{-1}=0$, $t_{l}=1$).

Suppose $f^{-1}(U_{i})$ is the  union of two disjoint open sets $V_{0}$ and
$V_{1}$. Because every fibre of $f$ is connected, we have
$f^{-1}(f(V_{j})) = V_{j}$ ($j=0,1$), and
the images $f(V_{0})$ and
$f(V_{1})$ are disjoint. As $f$ is a closed map, the image of the complement
of $V_{j}$ is closed. This is equal to the complement of $f(V_{j})$. Hence
$f(V_{j})$ is open.
 Then $f(V_{0})$ or $f(V_{1})$
must be empty, as $U_{i}$ is connected.
This implies that $f^{-1}(U_{i})$ is connected.

Take a point $c_{i}$ in each $f^{-1}(U_{i}\cap U_{i+1})$ for $1\le i\le l$
($U_{l+1}=U_{l}$)
and a path from $c_{i-1}$ to $c_{i}$ in $f^{-1}(U_{i})$ ($c_{0}=c_{l}$).
Define the closed path $\psi$ in $E$ by composing these paths. Then,
as every $U_{i}$ is contractible,
 $f\circ \psi$ is homotopic to $\varphi$.
\qed
\enddemo

\demo {\it Proof of lemma 3.3}
Suppose there is a sequence   $\{ c_{i}\}$ in $\HH_{d}(Z_{m})$ such that
$\Pi_{\pm}(c_{i})$ converges to an element $b \in
\Harm^{\pm}_{d}(S^{2m})$.
We shall show that $\{ c_{i}\}$ has a convergent subsequence.

Using  Pl{\"u}cker coordinates we fix an embedding of $Z_{m}$ into a large
complex projective space ${\bold C}P^{N}$. Each $c_{i}$ is represented as
$[F_{i,0};\dots;F_{i,N}]$ where the $F_{i,j}$ are polynomials such that, for
each $i$, $F_{i,0},\dots,F_{i,N}$ have
 no non-trivial common factor. For each $i$, the maximal degree of the
$F_{i,j}$'s is equal to $dD$, where $D$ is the degree of the embedding.
The coefficients of the polynomials give an embedding
$\{ c_{i} \} \longrightarrow {\bold C}P^{(dD+1)(N+1)-1}$.
We may take a subsequence of $\{ c_{i}\}$,  denoted also by $\{ c_{i}\}$,
such that $\{ c_{i}\}$ converges to some
$[F_{0};\dots;F_{N}]$ in the compact set
${\bold C}P^{(dD+1)(N+1)-1}$. Here the $F_{j}$'s may have a common factor.
Let $Q$ be the monic greatest common factor of the $F_{j}$'s and write
$F_{j}=QG_{j}$. Then $c=[G_{0};\dots;G_{N}]$ represents a horizontal
holomorphic map
to $Z_{m}$. Let $d'$ be the degree of $c$ as a map from $S^{2}$ to $Z_{m}$.
Then its degree as a map from $S^{2}$ to ${\bold C}P^{N}$ is $d'D$
and this is the maximal degree of the $G_{i}$'s. So $\deg\,Q=(d-d')D$.
Now $\{ c_{i}\}$ converges to $c$ on the complement of the roots of $Q$
and by assumption $\Pi_{\pm}(c_{i})$ converges to $b$ on $S^{2}$.
So $\Pi_{\pm}(c)$ and $b$ are harmonic maps which
agree on the complement of the roots of $Q$, hence on the whole of
$S^{2}$ by continuity. In particular, $d'=d$. This
implies that $Q=1$ and so $\{ c_{i}\}$ converges to $c$ on $S^{2}$.
\qed
\enddemo

$${}$$
\heading
\S 4. An explicit construction of full harmonic maps $S^2\longrightarrow
S^4$
\endheading

In this section we give an explicit construction of certain full harmonic
maps $S^2\longrightarrow S^4$. As a corollary of this (see proposition 4.2)
we obtain the following lemma:

\proclaim{Lemma 4.1}
If $d\ne 2$ then there is an element $f\in\HH_d(\C P^3)$ whose image
intersects the two-sphere $Z^+$ in a single point (with multiplicity
$1$).\qed
\endproclaim

\noindent This is equivalent to lemma 2.4, which was needed in the proof of
our main theorem. (It is obvious that lemma 2.4 implies lemma 4.1.
Conversely, given $f=[f_0;f_1;f_2;f_3]$ as in lemma 4.1, we may assume
without loss of generality that the intersection point is given by $z=0$.
Hence either
$[f_0;f_1;f_2;f_3]$ or $[f_2;f_3;f_0;f_1]$ satisfies the conditions of
lemma 2.4, as it is not possible to have $f_1(0)=f_3(0)=0$.)

The condition of lemma 4.1 can be explained in terms of the deformation in
\S 2 induced by the gradient flow of the Morse-Bott function $f^Q$ on $\CP$.
Suppose that the image of $f \in \hcp$ intersects $Z^+$ at $k$ points
(counted with multiplicity), i.e. that $f_0$ and $f_2$ have precisely $k$
common roots. The gradient flow deforms $f$ to a map
$D(f)\in\HH_{d-k}(Z^-)=\Hol_{d-k}(Z^-)$,
with $k$ {\it bubbles}. Let $q$ be the monic greatest common
factor of $f_0$ and $f_2$ and write $f_0 = qg_0$, $f_2 = qg_2$. Then
$D(f)=g = [g_0 ; g_2]$ and $\deg\,q=k$.

If $d \ge 2$, the condition of lemma 4.1 implies that $f$ is a full map. To
see this, suppose $f$ is non-full and that its image intersects $Z^+$ in one
point. As $f$ is a $d-$fold branched covering onto its image, the
multiplicity of the intersection is a multiple of $d$, hence larger
than one.  This is a contradiction. (There is no full map of degree 2, by
theorem 1.1(3), and this is
the reason why the lemma does not hold for $d = 2$.)

When $d>3$, the existence of a map satisfying the condition of the
lemma turns out to be almost equivalent to the existence of a
full map of degree $d$, a fact which is already known from Calabi's theorem.
We shall explain this in the context of our idea of considering deformations
associated to the Morse-Bott function on $\CP$.
Roughly speaking, we shall give a description of nearly all full maps which
intersect $Z^+$ simply at every intersection point, i.e., which converge to
maps into $Z^-$ with only simple bubbles. The lemma is an immediate
corollary of the construction, which we state as proposition 4.2 below.

One advantage of this alternative proof of the lemma is
that it may be helpful in investigating further the structure of $\hcp$
through Morse theory. However, we shall not pursue that direction in this
paper.

 We start from $g = [g_0 ; 0 ; g_2 ; 0] \in
\Hol_{d^\prime}(Z^-)\cong\Hol_{d^\prime}(S^2)$, with
$d^\prime=\deg\,g\ge2$. Assume that $\infty \in S^2$
is not a ramification point of $g$.
The degree of $\G$ is then precisely equal to $2d^\prime -2$
For simplicity, we assume also
that the ramification divisor of $g$ is the sum of $2d^\prime  -2$ distinct
points, i.e., $\G$ has only simple roots.
Let $q$ be a monic polynomial factor of $\G$
of degree $k$, with $1 \le k \le 2d^\prime  -2$.
Let $\A$ be the roots of $q$,  and let $\BT$ be the roots of
$(\G)/q$, where $N = 2d^\prime  -2-k$. We shall describe the maps $\ff \in
\hcp$ which converge to $g$ with bubbles at $\A$,
i.e.,  with $f_{0}=qg_{0}$ and $f_{2}=qg_{2}$.
The degree $d$ of $f$ is then equal to $d= d^\prime  +k$.

We shall give an expression for $f_1$ and $f_3$ in terms of $g_0$, $g_2$,
$q$
and a polynomial $h\in\Cal H(g,q)$, where $\Cal H(g,q)$ is the vector space
consisting of all polynomials $h$ of degree less than or equal to
$2d^\prime$, which satisfy the equations
$$
\pmatrix \format \r& \quad \r    \\
      g^\prime  _2(\be_j)  &  -g_2(\be_j)\\
     -g^\prime  _0(\be_j)  &   g_0(\be_j)\\
\endpmatrix
\pmatrix
h(\be_j)\\
h^\prime (\be_j)/2
\endpmatrix
=
\pmatrix
0\\
0\\
\endpmatrix ,\ \ j=1,\dots,N.
$$
Since the determinants of these matrices
are zero, the number of linearly independent equations for $h$ is
less than or equal to $N$, hence

$$\dim {\Cal H}(g,q) \ge 2d^\prime +1-N = k+3.$$
The significance of the definition of ${\Cal H}(g,q)$ may be explained as
follows. For any polynomial $h$, define $f_{1}$ and $f_{3}$ by
$$
\F. \tag*
$$
Then the functions $f_1$ and $f_3$ are polynomials
if and only if $h\in {\Cal H}(g,q)$.

Suppose this condition holds. Notice that the coefficients of
$z^{3d^\prime -1}$ in $g_2^\prime  h - g_2
h^\prime /2
$ and
$-g_0^\prime  h + g_0
h^\prime /2
$ are zero, hence
$$
\deg f_1, \deg f_3 \le k-( 2d^\prime -2) +3d^\prime -2 = d^\prime +k =d.
$$
Now, $\deg\,g=d'$ and $\deg\,q=k$, so the degree of $\ffq$ is equal to
$d=d^\prime +k$,
providing $qg_{0}$, $f_1$, $qg_{2}$ and  $f_3$ do not have
a common root. We shall prove that this is the case for a generic choice of
$g$ and $h$, and that the map $f$ is then a full horizontal holomorphic map.

\proclaim{Proposition 4.2}
Let $g = [g_0 ; 0 ; g_2 ; 0] \in \Hol_{d^\prime  }(Z^-)$  be a holomorphic map
from $S^2$ to $Z^-$ of degree $d^\prime   \ge 2$.
Let $q$ be a monic polynomial factor of $\G$
of degree $k$, with $1 \le k \le 2d^\prime  -2$.
For a generic $g$ and a generic $h \in {\Cal H}(g,q)$, the map
$$
\ffq \tag**
$$
is a full horizontal holomorphic map $S^2 \longrightarrow \CP$
of degree $d=d^\prime +k$, where $f_1$ and $f_3$ are defined by (*).
\endproclaim

Conversely, we shall prove:

\proclaim{Proposition 4.3}
Let $g = [g_0 ; 0 ; g_2 ; 0]$ be a holomorphic map
from $S^2$ to $Z^-$ of degree $d^\prime   \ge 2$.
Suppose that $\G$ is a polynomial of degree $2d^\prime -2$ with only
simple roots.
Then the construction of proposition 4.2 gives rise to every full horizontal
holomorphic map of degree $d=d^\prime +k$ of the form (**)
such that $q$ is a polynomial of degree $k\ge 1$ with only simple roots.
\endproclaim

\demo{Remark}
In proposition 4.3,
it is a consequence of the assumptions that $q$ is a factor of $\G$, i.e.,
that the bubbles appear only at ramification points of $g$.
\enddemo

\demo{Proof of proposition 4.3}
Let $f$ be a full horizontal holomorphic map of the form (**).
The horizontality equation
$$
f_0f_1^\prime  - f_0^\prime f_1 + f_2f_3^\prime  - f_2^\prime f_3 = 0
$$
is equivalent to:
$$
\gather
0  =  g_0^\prime  (\frac{f_1}{q}) - g_0(\frac{f_1}{q})^\prime
+ g_2^\prime  (\frac{f_3}{q}) - g_2(\frac{f_3}{q})^\prime  \\
{} =  (g_0\frac{f_1}{q} + g_2\frac{f_3}{q})^\prime
- 2(g_0^\prime \frac{f_1}{q} + g_2^\prime \frac{f_3}{q}), \\
\endgather
$$
In particular,
$$
q(\dfrac{g_0f_1 + g_2f_3}{q})^\prime
= 2(g_0^\prime f_1 + g_2^\prime f_3)
$$
is a polynomial, which implies
that $q$ is a factor of $q^\prime (g_0f_1 + g_2f_3)$.
We are assuming that $q$ has only simple roots, so $q$ and $q^\prime $ have
no non-trivial common factor, hence
$$
h= \dfrac{g_0f_1 + g_2f_3}{q}
$$
is a polynomial of degree $\le 2d^\prime$.
We rewrite  these equations as
$$
\pmatrix
      g_0  &  g_2\\
      g_0^\prime   & g_2^\prime  \\
\endpmatrix
\pmatrix
f_1 \\
f_3 \\
\endpmatrix
= q
\pmatrix
h \\
h^\prime /2 \\
\endpmatrix,
$$
which is equivalent to:
$$
\F.
$$
Because $qg_0$, $f_1$, $qg_2$ and $f_3$ have no common root, $q$ must
be a factor of $\G$. The condition for $f_1$ and $f_3$ to be polynomials is
given by
$h \in {\Cal H}(g,q)$ when $\G$ has only simple roots.
\qed
\enddemo

To prove proposition 4.2, we use the following lemma.

\proclaim{Lemma 4.4}
 For a generic $g$, and for any monic factor $q$ of $\G$ of degree $k$,  the
dimension of ${\Cal H}(g,q)$ is $k+3$.
\endproclaim

\demo{Proof}
Because  $\Hol_{d^\prime  }(Z^-)$ is an irreducible algebraic
variety, it suffices to give an example of $g\in \Hol_{d^\prime  }(Z^-)$
for which $\dim {\Cal H}(g,q) \le k+3$ for any $q$. (We already know that
$\dim {\Cal H}(g,q) \ge k+3$.)
Let $g_0 = 1$ and let $g_2$  be a polynomial such that $g_2^\prime$ has
only simple roots. Then the roots of $\G = -g_2^\prime  $
are all simple and the equations for $h$ include the equations
$$
h^\prime  (\be_j) = 0,  j = 1, \dots, N.
$$
These are linearly independent, hence for any $q$ which is a factor of $\G$,
the dimension of ${\Cal H}(g,q)$ is equal to
$2d^\prime +1 -N =k+3$.
\qed
\enddemo

\demo{Proof of proposition 4.2}
 We show that if $g$ is generic, then
there are finitely many proper linear subspaces of $\Cal H(g,q)$
such that if $h \in {\Cal H}(g,q)$ does not
lie on any of them, then the corresponding map
$$
\ffq
$$
is a full horizontal map of degree $d$.

It is easy to check that $f$ is horizontal because from the definition of
$f_1$ and $f_{3}$ we have
$$
\pmatrix
      g_0  &  g_2\\
      g_0^\prime   & g_2^\prime  \\
\endpmatrix
\pmatrix
f_1 /q \\
f_3 /q \\
\endpmatrix
=
\pmatrix
h \\
h^\prime /2\\
\endpmatrix.
$$
and, by eliminating $h$,
$$
g_0^\prime  (\frac{f_1}{q}) - g_0(\frac{f_1}{q})^\prime
+ g_2^\prime  (\frac{f_3}{q}) - g_2(\frac{f_3}{q})^\prime   = 0,
$$
which implies that the map
$$
\ffq = [g_0 ; \frac{f_1}{q} ; g_2 ; \frac{f_3}{q}]
$$
is horizontal.

To show that $f$ is a map of degree $d$, we have to show that $f_1$, $f_3$
and $q$ have no common root, for a generic $h$. Suppose that their greatest
common monic factor $w$ is not $1$,
and write $f_1 = w\tilde{f_1}$, $f_3 = w\tilde{f_3}$ and $q = w\tilde{q}$.
{}From proposition 4.3, there exists some  $\tilde{h}\in {\Cal H}(g,\tilde{q})$
from which we obtain the full map
$
[\tilde{q}g_0 ; \tilde{f_1} ; \tilde{q}g_2 ; \tilde{f_3}].
$
Notice that
$$
h = \dfrac{g_0f_1 + g_2f_3}{q}
  = \dfrac{g_0\tilde{f_1} + g_2\tilde{f_3}}{\tilde{q}} = \tilde{h},
$$
hence $h$ lies in the subspace ${\Cal H}(g,\tilde{q})$ of ${\Cal H}(g,q)$.
The dimension of ${\Cal H}(g,\tilde{q})$ is $\deg\ \tilde{q}+3$, which is
strictly
less than $\dim\ {\Cal H}(g,q)=k+3$. Here we have used lemma 4.4 twice.
Hence ${\Cal H}(g,\tilde{q})$ is a proper subspace of ${\Cal H}(g,q)$.

Since there are only finitely many
possibilities  for $\tilde{q}$, which is a monic factor of $q$ of degree less
than $k$,
a generic $h$ does not lie in the union of the
subspaces of the form ${\Cal H}(g,\tilde{q})$.
For such $h$, the polynomials $f_{1}$, $f_{3}$ and $q$ do not have a common
root.

Finally, we show that $\ffq$ is a full map.  If $f$ were not full, then the
degree of
$q$ would be a multiple of $d=d^\prime +k \ge 3$, by an argument used at
the beginning of this section. This is possible only if
$qg_0$ and $qg_2$ are identically zero, which is a contradiction.
\qed
\enddemo

\noindent{\it
Department of Mathematics, College of Arts and Sciences,
University of Tokyo,
\newline Meguro, Tokyo 153, Japan

\noindent
Department of Mathematics, University of Rochester,
\newline Rochester, NY 14627, USA

\noindent
Department of Mathematics, Faculty of Science, Toho University,
\newline Funabashi, Chiba 274, Japan

\noindent
Department of Mathematics, Tokyo Metropolitan University,
\newline Minami Ohsawa 1-1, Hachioji-shi, Tokyo 192-03, Japan}

\enddocument
\end